\documentclass[12pt]{article}
\usepackage{html}
\usepackage{epsf}
\usepackage{graphicx}
\usepackage{amssymb}
\newcommand{\araa}{Annual Review of Astronomy \& Astrophysics}
\newcommand{\apj}{The Astrophysical Journal}
\newcommand{\prd}{The Physical Review D}
\newcommand{\aap}{Astronomy \& Astrophysics}
\newcommand{\nar}{New Astronomy Reviews}
\newcommand{\planck}{\emph{Planck}}
\newcommand{\lcdm}{$\Lambda$CDM}
\begin{document}
\title{MATTERS OF GRAVITY, The newsletter of the APS Topical Group on 
Gravitation}
\begin{center}
{ \Large {\bf MATTERS OF GRAVITY}}\\ 
\bigskip
\hrule
\medskip
{The newsletter of the Topical Group on Gravitation of the American Physical 
Society}\\
\medskip
{\bf Number 42 \hfill Fall 2013}
\end{center}
\begin{flushleft}
\tableofcontents
\vfill\eject
\section*{\noindent  Editor\hfill}
David Garfinkle\\
\smallskip
Department of Physics
Oakland University
Rochester, MI 48309\\
Phone: (248) 370-3411\\
Internet: 
\htmladdnormallink{\protect {\tt{garfinkl-at-oakland.edu}}}
{mailto:garfinkl@oakland.edu}\\
WWW: \htmladdnormallink
{\protect {\tt{http://www.oakland.edu/?id=10223\&sid=249\#garfinkle}}}
{http://www.oakland.edu/?id=10223&sid=249\#garfinkle}\\

\section*{\noindent  Associate Editor\hfill}
Greg Comer\\
\smallskip
Department of Physics and Center for Fluids at All Scales,\\
St. Louis University,
St. Louis, MO 63103\\
Phone: (314) 977-8432\\
Internet:
\htmladdnormallink{\protect {\tt{comergl-at-slu.edu}}}
{mailto:comergl@slu.edu}\\
WWW: \htmladdnormallink{\protect {\tt{http://www.slu.edu/colleges/AS/physics/profs/comer.html}}}
{http://www.slu.edu//colleges/AS/physics/profs/comer.html}\\
\bigskip
\hfill ISSN: 1527-3431

\bigskip

DISCLAIMER: The opinions expressed in the articles of this newsletter represent
the views of the authors and are not necessarily the views of APS.
The articles in this newsletter are not peer reviewed.

\begin{rawhtml}
<P>
<BR><HR><P>
\end{rawhtml}
\end{flushleft}
\pagebreak
\section*{Editorial}

The next newsletter is due February 1st.  This and all subsequent
issues will be available on the web at
\htmladdnormallink 
{\protect {\tt {https://files.oakland.edu/users/garfinkl/web/mog/}}}
{https://files.oakland.edu/users/garfinkl/web/mog/} 
All issues before number {\bf 28} are available at
\htmladdnormallink {\protect {\tt {http://www.phys.lsu.edu/mog}}}
{http://www.phys.lsu.edu/mog}

Any ideas for topics
that should be covered by the newsletter, should be emailed to me, or 
Greg Comer, or
the relevant correspondent.  Any comments/questions/complaints
about the newsletter should be emailed to me.

A hardcopy of the newsletter is distributed free of charge to the
members of the APS Topical Group on Gravitation upon request (the
default distribution form is via the web) to the secretary of the
Topical Group.  It is considered a lack of etiquette to ask me to mail
you hard copies of the newsletter unless you have exhausted all your
resources to get your copy otherwise.

\hfill David Garfinkle 

\bigbreak

\vspace{-0.8cm}
\parskip=0pt
\section*{Correspondents of Matters of Gravity}
\begin{itemize}
\setlength{\itemsep}{-5pt}
\setlength{\parsep}{0pt}
\item Daniel Holz: Relativistic Astrophysics,
\item Bei-Lok Hu: Quantum Cosmology and Related Topics
\item Veronika Hubeny: String Theory
\item Pedro Marronetti: News from NSF
\item Luis Lehner: Numerical Relativity
\item Jim Isenberg: Mathematical Relativity
\item Katherine Freese: Cosmology
\item Lee Smolin: Quantum Gravity
\item Cliff Will: Confrontation of Theory with Experiment
\item Peter Bender: Space Experiments
\item Jens Gundlach: Laboratory Experiments
\item Warren Johnson: Resonant Mass Gravitational Wave Detectors
\item David Shoemaker: LIGO Project
\item Stan Whitcomb: Gravitational Wave detection
\item Peter Saulson and Jorge Pullin: former editors, correspondents at large.
\end{itemize}
\section*{Topical Group in Gravitation (GGR) Authorities}
Chair: Daniel Holz; Chair-Elect: 
Beverly Berger; Vice-Chair: Deirdre Shoemaker. 
Secretary-Treasurer: James Isenberg; Past Chair:  Manuela Campanelli;
Members-at-large:
Michael Landry, Nicolas Yunes,
Curt Cutler, Christian Ott,
Kimberly Boddy, Benjamin Farr,
Andrea Lommen, Jocelyn Read.
\parskip=10pt

\vfill
\eject

\vfill\eject

\section*{\centerline
{we hear that \dots}}
\addtocontents{toc}{\protect\medskip}
\addtocontents{toc}{\bf GGR News:}
\addcontentsline{toc}{subsubsection}{
\it we hear that \dots , by David Garfinkle}
\parskip=3pt
\begin{center}
David Garfinkle, Oakland University
\htmladdnormallink{garfinkl-at-oakland.edu}
{mailto:garfinkl@oakland.edu}
\end{center}

Deirdre Shoemaker was elected Vice Chair of GGR; 
Jocelyn Read and Andrea Lommen were elected Members at large of the Executive Committee of GGR.
Kimberly Boddy was elected Student Representative of GGR.

Lisa Barsotti has won the IUPAP General Relativity and Gravitation Young Scientist Prize.  Aseem Paranjape has won
the J\"{u}rgen Ehlers Thesis Prize.  Aron Wall has won the Bergmann-Wheeler Thesis Prize.  Paul Fulda has won the GWIC Thesis prize. 

Yvonne Choquet-Bruhat, Natalie Deruelle, George Ellis, Eanna Flanagan, Stefan Hollands, Werner Israel, Bala Iyer, Luis Lehner, 
Ted Newman, Peter Saulson, Bernard Schutz, Masaru Shibata, Tarun Souradeep, Takahiro Tanaka, and Robert Wald have been 
been elected Fellows of the International Society for General Relativity and Gravitation.  

The following people have been elected to the following offices in the International Society for General Relativity and Gravitation: Gary Horowitz as President; Beverly Berger as Secretary; Jiri Bicak, Stefano Liberati, Takashi Nakamura, Miguel Alcubierre, Jorma Louko, B. Sathyaprakash, Eanna Flanagan, and Don Marolf as members of the International Committee; Nathalie Deruelle and Luis Lehner as Auditors; Emanuele Berti, Patrick Brady, Brien Nolan, Masaru Shibata,  and Alicia M. Sintes Olives, 
as members of the Nominating Committee.

Majd Abdelqader, Valentina Baccetti, Christopher Berry, Sydney Chamberlin, Stephanie Erickson, Oliver Gerberding, Lisa Glaser, Gavin Hartnett, Giuliana Russano, and Antonia Zipfel have been awarded the Hartle prize for best graduate student presentations at GR20/Amaldi10.  Alexandre Le Tiec, Charles Melby-Thompson, Eric Perlmutter, and John Veitch have been awarded the Chandrasekhar prize for best postdoctoral presentations at GR20/Amaldi10.

Hearty Congratulations!

\vfill\eject
\section*{\centerline
{What's new in LIGO}}
\addtocontents{toc}{\protect\medskip}
\addcontentsline{toc}{subsubsection}{
\it What's new in LIGO, by David Shoemaker}
\parskip=3pt
\begin{center}
David Shoemaker, MIT
\htmladdnormallink{dhs-at-mit.edu}
{mailto:dhs@mit.edu}
\end{center}

Here is a brief update on the advances in LIGO -- our name for the
LIGO Laboratory (Caltech/MIT) and the greater LIGO Scientific
Collaboration (LSC). To date, all observations have been made with `first generation detectors' -- initial LIGO, in collaboration with the European Virgo and GEO detectors. Looking forward, the principal focus of the Collaboration at this time is on Advanced LIGO -- the instruments, the analysis pipelines, and the astrophysics we can extract from the anticipated gravitational waveforms.

\medskip
{\bf Astrophysical interpretation} of the observing runs undertaken with initial/enhanced LIGO/GEO jointly with the Virgo detector has led to a number of interesting upper limits for gravitational waves as well as some `non-detections' which exclude some potential astrophysical models. We mention some high points.

- GRB070201 was a short hard gamma-ray burst observed in the x-ray band by satellites 
in the Interplanetary Network.  The error box had significant overlap with the M31 (Andromeda) galaxy located 730 kpc from the Milky Way galaxy, thus making it a prime target for gravitational-wave searches. A binary 
merger would produce a characteristic gravitational-wave signal that would have been  easily detectable at the distance of M31. Data from two LIGO Hanford detectors were analyzed in a narrow time window around GRB070201 using both template-based searches for binary mergers and burst search algorithms (described below).  No signal was found, thus excluding the hypothesis of a compact binary (either a binary neutron star or black hole-neutron star) merger as the progenitor of GRB070201 located in M31 at 90\% and 99\% 
confidence levels, respectively.

- A key threshold for pulsars is the spin-down limit, the value of gravitational-wave strain for which the observed rotational braking would be entirely accounted for by gravitational-wave emission.  Using LIGO and Virgo data, we can set a limit on gravitational wave emission that is lower than the spin down limit for the Crab and Vela pulsars.  
The Crab pulsar is particularly appealing because it is relatively young and rapidly spinning-down.  
While the predominant energy dissipation mechanisms are likely to be magnetic dipole radiation or charged particle emission in the pulsar's magnetosphere, the measured braking index of the Crab pulsar suggests that neither dipole 
radiation nor particle ejection can account entirely for the rotational slow down. Using a subset of data from LIGO's S5 science run, the LIGO Scientific Collaboration searched for gravitational wave emission from the Crab pulsar during a nine 
month duration during which no pulsar timing jumps occurred.   No gravitational waves were observed, and the data was used to set upper limits on the strain $h \le 3.3  \times 10^{-25}$ and ellipticity $\epsilon \le 1.79 \times 10^{-4}$. Further analysis leads to a limit on gravitational wave strain corresponding to no more than 2\% of the energy emitted by the Crab pulsar being in the form of gravitational waves.  

- By cross-correlating S5  data from the LIGO Livingston and Hanford 4 km interferometers, an upper limit of $\Omega_{GW} \le 6.9  \times 10^{-6}$ has been established on the 
energy density of stochastic gravitational waves (normalized to the closure energy density 
of the universe) assuming the gravitational wave background is confined within the 50-150 Hz frequency band.   This is the best experimental limit in the LIGO frequency band, beating the limit inferred from Big Bang 
Nucleosynthesis by almost a factor of 2.

\medskip

{\bf Advanced LIGO} is a replacement for the initial LIGO detectors. The NSF granted funds for the effort to fabricate, install, and bring to `lock' three interferometers; the UK, Germany, and Australia also made contributions of designs and hardware. The Project {\it per se} started in 2008 and will complete in 2015. As of August 2013 the fabrication is completed, and most hardware has been assembled, tested, and installed at the two LIGO Observatories (Hanford, Washington and Livingston, Louisiana). Testing of ever more complete subsets of the hardware is now underway, and by early 2014  instruments at both Observatories will be complete. The third interferometer is being put in storage with the plan that it can be installed in an infrastructure to be built in India, improving the pointing precision of the world-wide array of detectors. The Virgo Collaboration is also making improvements to its detector in Cascina, Italy, and a new detector in Japan - KAGRA - is in construction. 

{\bf The Advanced LIGO Sensitivity} once fully commissioned will be roughly ten times better than initial LIGO, and will extend to roughly 10 Hz in contrast to initial LIGO's 40 Hz. Estimates for the detection rates of neutron-star inspirals (http://arxiv.org/abs/1003.2480) are for roughly 40 detections per year for the ultimate sensitivity (with an order of magnitude uncertainty in rate in either direction). The instrument will be commissioned in phases interleaved with observing runs; the plan is to carry out an observing run at a sensitivity of a duration capable of a first detection in 2016. Gradually Advanced LIGO, and the other 'second generation' detectors, will reach their full sensitivity in the years to follow, allowing both more precision astrophysical interpretations and better pointing for joint electromagnetic and gravitational-wave 'multi-messenger' astrophysics.

\medskip
 {\bf Instrument Research} in the LIGO Scientific Collaboration is working toward both incremental improvements in the sensitivity of Advanced LIGO and similar instruments, and starting consideration of `third generation' instruments which could make another order of magnitude improvement in ground-based gravitational wave detectors.  The research can be broken into several key areas:
 
 - Thermal noise limits Advanced LIGO in the mid-band and more sophisticated instruments in a broad range of (lower) frequencies. Reductions in the noise could be gained by changes in the optics coatings (whose mechanical loss is now the dominant noise source in the mid band), extrapolations to larger test masses and longer suspension systems, or ultimately cryogenics to reduce the $k$T per mode. 
 
 - Quantum noise  limits Advanced LIGO at low frequencies (radiation pressure) and high frequency (shot noise), although the two are linked via the signal recycling topology. Near-term changes to add a source of squeezed vacuum light to the output port has already shown success in a trial on initial LIGO, and will be an early upgrade of Advanced LIGO -- probably with frequency-dependent squeezing. More exotic interferometer topologies and readout systems which allow more manipulation of the quantum state to improve sensitivity will be prototyped, and may be either part of Advanced LIGO upgrades or play a role in third-generation instruments.
 
 - Newtownian noise, or gravitational-gradient noise will just start to limit Advanced LIGO around 10-15 Hz. Methods to regress out the contribution using arrays of seismometers are in study, and will certainly play a role in future instruments.
 
 - Technical noise sources -- laser, seismic, environmental more generally -- are always important in these instruments and will be `brought along' to realize the performance possible with the more fundamental improvements in the detectors.
 
\medskip

{\bf In summary,} The work around the LIGO detectors has been quite lively in the past few years, and is on the cusp of having instruments which promise to make discoveries on a time scale that corresponds to a graduate student's research interval. We are all as excited as those graduate students!

\vfill\eject

\section*{\centerline
{News from NSF}}
\addtocontents{toc}{\protect\medskip}
\addcontentsline{toc}{subsubsection}{
\it News from NSF, by Pedro Marronetti}
\parskip=3pt
\begin{center}
Pedro Marronetti, NSF
\htmladdnormallink{pmarrone-at-nsf.gov}
{mailto:pmarrone@nsf.gov}
\end{center}

Fiscal year 2013 has been marked by a series of special circumstances that have taken their toll in the funding made available to all the research programs in the Division of Physics at NSF. The American Taxpayer Relief Act of 2012 (a.k.a. sequester) as well as the fact that the Government operated under a continuing resolution all throughout the year, resulted in an uncharacteristic delay in the confirmation of the appropriations for each Division of our Directorate. Physics Division budgets were handed out during the last days of July, when this usually happens in February. However, the most serious consequence was the overall reduction of the funds available for individual programs. The fiscal year 2012 NSF budget of \$7,105 million was reduced in 2013 to \$6,884 million (3.1\%). This resulted in a reduction of 4.5\% of the Directorate of Mathematical and Physical Sciences (MPS). Physics Division, one of the five Divisions of MPS, saw its overall budget decline from last year’s \$277.4 million to \$250.7 million (9.6\%). To complicate matters, MPS was mandated to keep the funding level for facilities unchanged, leaving the individual programs funding to bear the brunt of the cuts. Thus, Physics Division programs experienced a reduction of approximately 12\%.  For 2014, NSF has requested \$7,626 million; an increase of 10.8\% with respect of the 2013 budget. This includes an increase of 10.9\% for MPS and 15.3\% for Physics Division.

Last year we implemented a new Mentoring program directed at schooling investigators in the art of writing succesful NSF proposals in general and in the areas of Gravitational Physics in particular. Close to twenty experienced NSF-funded researchers have so far volunteered their time to review proposal drafts and provide some feedback to the submitters. The program is open to anybody submitting a proposal to any of the Gravitational Physics programs. This past year we received about ten mentoring requests and the comments I received have been very positive. If you are interested in participating in the program, either as a mentor or as mentee, send me an email (pmarrone@nsf.gov).

Finally, the Physics at the Information Frontier – Computational Physics program (PIF/CP) is now part of a program called Computation and Data Enable Science and Engineering (CDS\&E).  All proposals submitted to PIF/CP are automatically part of this initiative that aims at facilitating the co-reviewing and potential co-funding of the proposals by other Divisions. For more information on this program, please visit 
\htmladdnormallink
{\protect {\tt{http://www.nsf.gov/funding/pgm\_summ.jsp?pims\_id=6681\&org=PHY\&from=home}}}
{http://www.nsf.gov/funding/pgm\_summ.jsp?pims\_id=6681&org=PHY&from=home}\\
or contact Bogdan Mihaila, Program Director for PIF/CP (bmihaila@nsf.gov).

\vfill\eject
\section*{\centerline
{Citation counts and indices: Beware of bad data}}
\addtocontents{toc}{\protect\medskip}
\addcontentsline{toc}{subsubsection}{
\it Citation counts and indices: Beware of bad data, by Clifford Will}
\parskip=3pt
\begin{center}
Clifford Will, University of Florida
\htmladdnormallink{cmw-at-physics.ufl.edu}
{mailto:cmw@physics.ufl.edu}
\end{center}

In recent years, citations counts and associated indices have become increasingly important in a wide range of professional considerations, including hiring, promotions and grant reviews.  Citation counts can make the difference between a cut or an increase in funding a university department by a national or local government.  It is not uncommon to see post-doc applications listing citation counts for every paper on the candidate's CV.  
I have seen publication lists on CVs and webpages that are simply the output of a SPIRES search on the individual's name.   And if a paper should make one of SPIRES' ``famous'' or ``renowned'' paper categories, you can be sure that that fact will be indicated prominently somewhere on the CV.   

Citation counts are great for bureaucrats and administrators, because, by definition, they are quantitative.   They can be added, subtracted, normalized and plotted.  Indices based on these counts abound: the ubiquitous Hirsch or {\em h}-index (the first number $n$ such that in a ranking of papers by citation count, the rank of the $n+1$th paper exceeds the number of its citations) being the most prominent among many. 

Debate continues to rage as to whether these counts and indices actually mean anything.  Do they really measure the importance or impact of a researcher's work on the field?   How does one normalize for multiple author papers, especially where the number of authors can reach into the hundreds?  Should review articles count the same as original research articles?   How does one normalize across fields with different citation traditions?  Is one paper with a zillion citations better than a lot of papers with modest citations?  Do these indices favor experimentalists over theorists?
And so on.

But one issue that I have not seen debated is the accuracy of the actual citation data.  We take for granted that every citation of every paper we have written has been counted.

But this is not necessarily so.  Case in point:

In 2008 I wrote a paper for {\em Astrophysical Journal Letters} on the possibility of testing the black hole no-hair theorems using measurements of stars orbiting the galactic center black hole SgrA$^*$.   I thought it was a pretty good paper which seemed to have had some impact, and so I recently went to INSPIRE (the latest version of SPIRES) to check on the citations.   I was astonished to find that it had been cited only 13 times in 5 years.
On the other hand, according to the NASA/ADS database, it had been cited 73 times.  What happened to the other 60 citations?    Most of them were in standard journals like {\em Phys. Rev. D} and {\em Astrophys. J}.   I had assumed that, with modern DOI designations and search engines, everything would be caught, but apparently not. 

To correct this, I had to generate a list (via NASA/ADS) of all the missing citations and send them to SPIRES.  The staff there then entered the information by hand.  The list is now accurate (and my {\em h}-index went up by one!  But I'm not obsessed .... really, I'm not...) 

But this raises the question: What else might be missing?

In all fairness, SPIRES was set up as a database primarily for the high-energy physics community, and the SPIRES staff admit that they have difficulty getting all the references from the various astronomy and astrophysics journals.  In view of the increasing links between particle physics and astronomy, they told me that they are considering talking to the NASA/ADS people about ways of doing this better. 

But for us in gravitational physics, who publish in a wide range of journals not automatically associated with high-energy physics, a citation count provided by a system such as SPIRES might not be as accurate as you think.   So like every good physicist, you should first check the quality of the data.

\vfill\eject
\section*{\centerline
{An overview of the initial cosmological results from {\it Planck}}}
\addtocontents{toc}{\protect\medskip}
\addtocontents{toc}{\bf Research briefs:}
\addcontentsline{toc}{subsubsection}{
\it Results from Planck, by William Jones}
\parskip=3pt
\begin{center}
William Jones, Princeton University
\htmladdnormallink{wcjones-at-princeton.edu}
{mailto:wcjones@princeton.edu}
\end{center}

The cosmic microwave background radiation (CMB) is the relict thermal
radiation pervading all space with a characteristic temperature near
2.73 Kelvin~\cite{cmb_predict64,peebles65,dprw65}. Having been
generated during a hot and relatively dense evolutionary stage of the
Universe (before the formation of stars, galaxies or clusters of
galaxies), the statistical properties of anisotropies in the intensity
and polarization of the CMB provide a unique laboratory for the study
of the contents and evolution of the
Universe~\cite{cmbnobel78,hu02,cmbnobel06}.  In the context of a
leading class of early Universe theories, known collectively as
Inflation, the origin of these anisotropies is traced back to the
magnification of quantum fluctuations of an otherwise isotropic metric
(of both the scalar- and tensor-type) during an early epoch of nearly
exponential expansion of space.  To the extent that these
theories make testable predictions, the CMB can be used to probe
physics at energy scales that are beyond the reach of any terrestrial
laboratory.

 Since its discovery, experimental efforts have raced to exploit the
 scientific potential of the CMB.  During the course of the last
 twenty years, pioneering ground- and balloon-borne CMB anisotropy
 experiments provided an observational foundation for the now-standard
 \lq\lcdm\rq\ cosmological model, wherein \emph{all} cosmological
 observables can be described by six
 parameters~\cite{lange01,sask,toco,archeops,b2k}.

The \planck\ mission is the third space observatory dedicated to
measurements of the CMB, following on the hugely successful COBE and
WMAP missions~\cite{cobe93,mog6,mog32}.  The results from WMAP
dramatically improved upon the precision of the parameters describing
the previously established standard cosmology. The \planck\ data have
the statistical power not only to further refine our parameter
estimates, but also to quantify the support for extensions to the
six-parameter cosmology and to discriminate between competing theories
of the early Universe.

\planck\ has produced all-sky surveys of continuum emission
in nine frequency bands from 30 GHz to 1 THz, with angular resolution
ranging from 30 to 5 arc-minutes.  All but the two highest frequency
bands are sensitive not only to the total intensity, but also to the
linear polarization of the emission.  Launched on May 14, 2009,
\planck\rq s scientific survey began on August 12, 2009. The High
Frequency Instrument (HFI) completed its final observations on January
16, 2012.

\begin{figure}
\begin{center}
\includegraphics[width=0.6\textwidth]{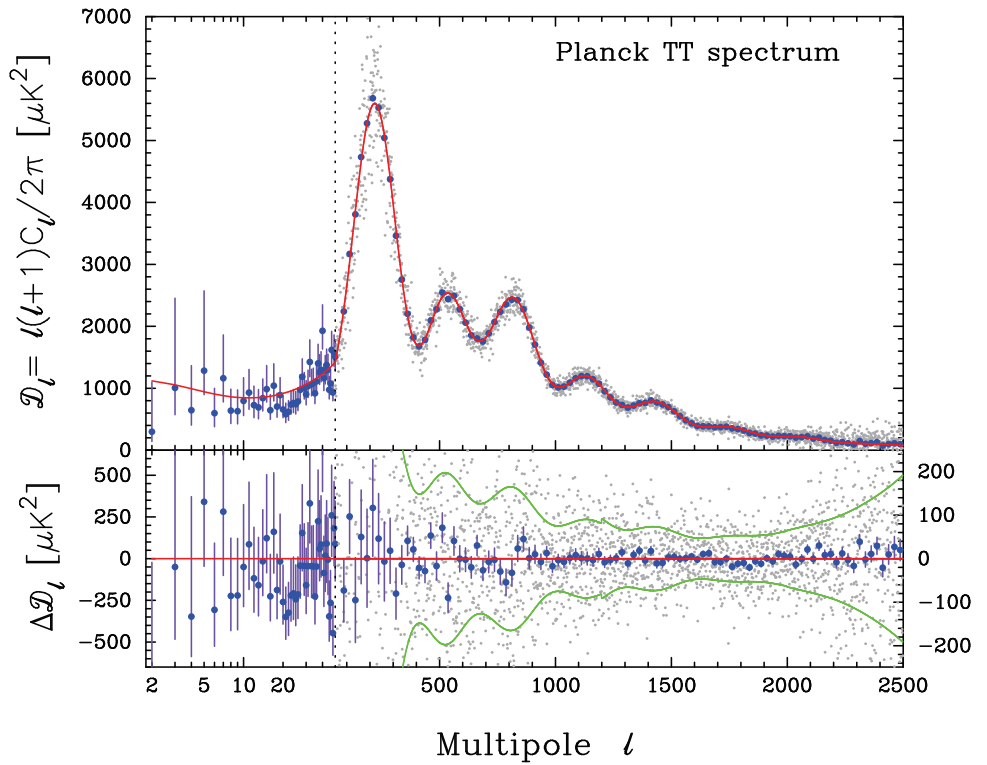}
\caption{The angular power spectrum of the the CMB anisotropies
  measured by \planck.  The vertical line delimits the transition from
  a log to linear scale on the multipole moment, which is inversely
  proportional to the angular size of features in the CMB.  The upper
  panel shows the un-binned power spectrum (grey) along with the data
  binned with $\Delta\ell\simeq 31$. The error bars on the binned
  estimates include the contribution of sample variance, statistical
  noise and foreground modeling. The best fit \lcdm\ cosmology is
  shown in red.  The lower panel shows the residuals of the data with
  respect to the best fit model.  Note the change in the
  \emph{vertical} scale at $\ell = 50$.  The light green line shows
  the envelope derived from the diagonal of the full covariance of the
  un-binned power spectrum.  The excursions evident on the largest
  scales ($\ell \lesssim 50$) are highly robust in the \planck\ data,
  and are consistent with those observed by WMAP.\label{fig:plancktt}}
\end{center}
\end{figure}

The \planck\ data represent the first ever all-sky surveys between 100
GHz and a THz, providing a component map of the CMB with
noise more than an order of magnitude lower than previously available,
a qualitatively new understanding of diffuse Galactic
emission as well as catalogs of Galactic and extra-Galactic compact
sources.  Here we summarize the cosmological results
from \planck\ derived from the first 15.5 months of intensity data. A
survey of the full breadth of astrophysical results, embodied in 29
papers released on March 21, 2013, can be found in the mission summary
paper~\cite{planck13}.  The analysis of the full data set, including the
polarization results, will be published in the first half of 2014.

\vspace{3mm}

\noindent {\bf \planck\ Cosmology}

The \planck\ data provide an extremely strong confirmation of the
six-parameter \lcdm\ paradigm.  A simple model of a geometrically flat
Universe consisting of a cosmological constant, dark matter and trace
amounts of baryonic matter is sufficient to account for \emph{all}
cosmological observations to date.  Independent measurements of the
CMB, primordial nucleosynthesis (BBN), the large scale distribution of
galaxies are jointly and separately consistent with the same
underlying Cosmology.  

\begin{figure}
\begin{center}
\includegraphics[angle=270,width=0.6\textwidth]{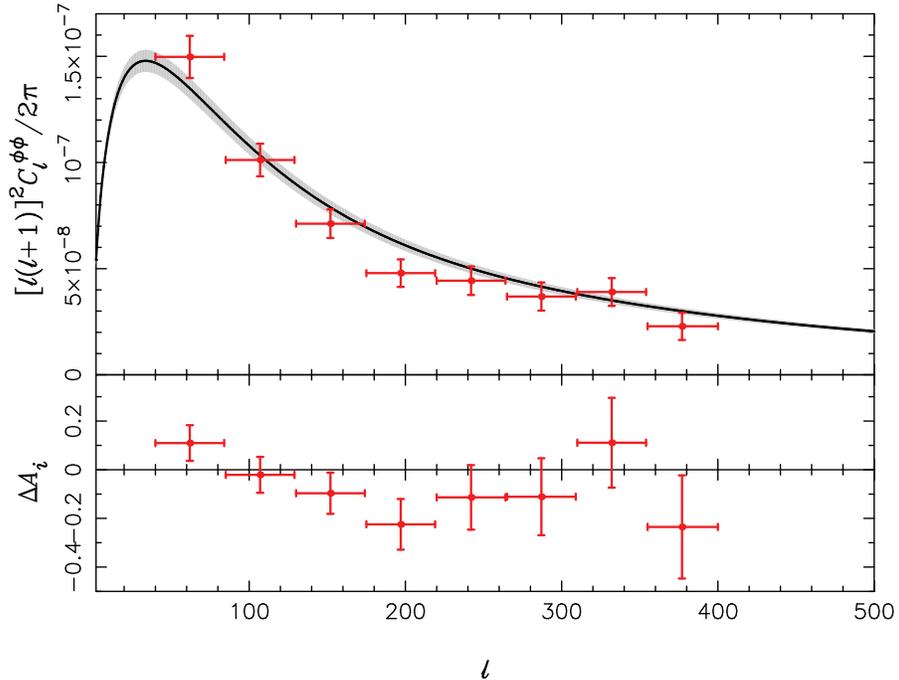}
\vskip0.5truein
\caption{The lensing power spectrum derived from the 143 and 217 GHz
  \planck\ data.  The upper panel shows the power spectrum
  $\mathcal{C}_{\ell}^{\phi\phi}$, derived from the trispectrum of the
  \planck\ maps.  The black line and grey envelope represent the
  result \emph{predicted} from the best fit cosmology derived from the
  CMB angular power spectrum, subject to the uncertainties in the
  \lcdm\ parameters.  From the point of view of parameter estimation,
  the two represent essentially independent
  constraints.\label{fig:lens}}
\end{center}
\end{figure}

The phenomenological power of the \lcdm\ model is evident in Figure
\ref{fig:plancktt}.  The upper panel shows the power spectrum of the
intensity fluctuations measured by \planck, with the best-fit
\lcdm\ model prediction over plotted in red, while the lower panel
shows the residuals between the data and the model.  The measurement
of the power spectrum of the CMB fluctuations is limited by sample
variance on angular scales above $\sim 10^\prime$ and by a combination
of sample variance and foreground confusion on smaller
scales\footnote{The sample variance limit is a result of the fact that
  we only have one Universe to observe, and it is obscured by our
  Galaxy.}.

Although the parameterization used in the CMB analysis differs in
order to minimize the impact of degeneracies in the power spectrum,
the six physical parameters can be understood in terms of the
matter-energy content (a cosmological constant, dark matter and
baryons), the initial conditions (an amplitude and spectral index of
the spatial power spectrum of primordial scalar fluctuations) and a
single parameter describing the late time effect of the ionizing
radiation from the first population of stars (the optical depth to
reionization).  Our Universe is presently dominated by a combination
of a cosmological constant (roughly 68\% of the total) and dark matter
(roughly 27\%).  The bulk of the remainder (just about 5\%) consists
of hydrogen and helium, with trace amounts of the heavier elements and
neutrinos\footnote{For further discussion of the methodology and
  parameter constraints, I will refer the reader not only to the
  \planck\ publications, but also to the excellent summary of the
  standard model parameters provided in Lyman Page\rq s submission to
  the thirty-second issue of Matters of Gravity~\cite{planck_cosmo13,mog32}}.

The non-uniform distribution of matter along the line of sight to the
surface of last scattering results in the weak lensing of the
intrinsic CMB fluctuations, impacting both the two point function and
the higher order statistics of the CMB component map (see Figure
\ref{fig:lens}).  The \planck\ data provide a highly significant
statistical measurement of these effects, providing for the first time
a unique constraint on the matter content of the Universe.
Importantly, these CMB-derived constraints help to break the long
standing geometric degeneracy between the impact of a cosmological
constant and dark matter on the CMB power spectrum.  The lensing
provides an essentially independent limit on the matter content,
allowing constraints on the spatial curvature and the Hubble parameter
from the CMB alone.

Although the \lcdm\ model provides an excellent fit to the data, the
statistical power of the \planck\ data enable rigorous limits on
the support for extensions to the basic six-parameter model.  These
extensions include non-zero spatial curvature, neutrino physics,
non-standard primordial helium abundances, dark energy (as opposed to
a simple cosmological constant) and features in the primordial power
spectrum.  The consistency of the data with the standard cosmology
requires the existence of the three known light neutrino species and
the BBN prediction for the primordial helium fraction, given the
neutrino population and the physical baryon density.  The
\planck\ data put very stringent limits on each of these single
parameter extensions to the standard model, and in no case support
deviations from the standard model assumptions (see, for example,
Figure 21 of \cite{planck_cosmo13}).

\planck\rq s phenomenal sensitivity and dynamic range in angular scale
provide unprecedented sensitivity to the presence of isocurvature
admixtures and to non-Gaussianity, making the experiment a powerful
probe of early Universe theories.  The leading class of such theories,
referred to collectively as Inflation, posit an epoch of near
exponential expansion of space in the very early Universe.  Such
an event would provide many, if not all, of the initial conditions
needed to explain the observed uniformity, geometric flatness and
structure of the Universe~\footnote{Such a phenomenon is not without
  precedent; the data favoring a large cosmological constant suggest
  that we are entering a new stage of Inflation just now, however
  unlikely that may seem from a frequentist point of view.}.

The \planck\ measurement of broken scale invariance ($n_s < 1$) is
precisely that expected in the simplest Inflationary scenarios; the
requirement that the Inflationary epoch eventually ends imprints a
characteristic length scale in the primordial fluctuations manifests
itself in broken scale invariance. The \planck\ limits on the local
type of the primordial bispectrum (three-point function) are at the
level of the ISW-lensing bias, which are a result of the correlation
of the lensing potential and the (intrinsically Gaussian) secondary
CMB anisotropies. The \planck\ data are entirely consistent with the
predictions of single-field slow-roll Inflationary theories, and are
not consistent with early Universe theories involving isocurvature
admixtures or appreciable levels of non-Gaussianity in the CMB
anisotropies.

The robust detection of broken scale invariance, tight limits on the
level of non-Gaussianity, and new upper limits on the ratio of
primordial tensor- and scalar-type fluctuations in the CMB have begun
to observationally discriminate between competing early Universe
theories.


\vspace{3mm}
\noindent {\bf Cosmic Conundrums}
 
There are further tantalizing hints that new physics may be lurking in
the wings - quite literally the wings of a statistical distribution.
On the largest scales - corresponding to angular separations larger
than about eight degrees - the fluctuations measured by \planck\ (and
WMAP before her) are smaller than expected in the standard
model.  The effect is subtle but persistent, and since \planck\rq
s best fit cosmology predicts more power on these scales than
the previous estimates from WMAP, the statistical significance of the
effect detected by \planck\ is even higher than the WMAP estimate.

The standard cosmological model describes a Universe that is, at first
blush, incredibly simple; the basic six-parameter cosmology passes an
astounding battery of observational tests, each subject to completely
independent systematic errors and together sensitive to all known
physical laws applied over scales ranging from the nuclear
interactions to the size of the observable Universe.

And yet, despite the phenomenological success of the theory, we find
that it is ultimately a Ptolemaic view of the Universe - that we live
at a privileged time, and one that is seemingly highly tuned.  How can
we understand a cosmological constant (vacuum-energy) that is finely
tuned such that it has only just come to dominate the matter density?
How can we address the tuning and unlikeliness problems that plague
the Inflationary early Universe theories~\cite{steinhardt}?  An
understanding of Inflation, if not the current epoch then the one that
appears to have been responsible for the birth of our Universe, will
need to be confronted by the data from the next generation of
dedicated CMB polarization missions.

\vfill\eject
\section*{\centerline
{Gravity - New Perspectives from Strings and Higher Dimensions}}
\addtocontents{toc}{\protect\medskip}
\addtocontents{toc}{\bf Conference reports:}
\addcontentsline{toc}{subsubsection}{
\it Gravity - Strings and Higher Dimensions, 
by Simon Ross}
\parskip=3pt
\begin{center}
Simon Ross, Durham University 
\htmladdnormallink{S.F.Ross-at-durham.ac.uk}
{mailto:S.F.Ross@durham.ac.uk}
\end{center}

A workshop on ``Gravity - new perspectives from strings and higher dimensions'' was held in the lovely setting of the Centro de Ciencias de Benasque Perdo Pascual, in the Spanish Pyrenees, from the 14th to the 26th of July 2013. This was the third in a continuing series of workshops on gravity at the centre, and continued a strong tradition of combining interesting talks with lively discussion. 

The programme began with a marathon ($>$ 3 hour) talk by Don Marolf, who reviewed firewalls and other approaches to black hole information, giving a balanced view of this controversial subject, and gave new arguments for the genericity of firewalls. The related subject of fuzzballs and black hole microstates was discussed first in a lively informal discussion led by Iosif Bena, which focused on attempts to construct non-supersymmetric examples, and progress on constructing superstrata, which are expected to play a key role in accounting for the entropy of three-charge black holes. The next day, Nick Warner explained how the known microstate solutions evade theorems in GR forbidding the existence of smooth solitons without horizons in theories with massless matter. This work provides a useful new insight into the key features that lead to the smooth micro state solutions.

There were a large number of talks related to holography and AdS/CFT, relecting the continued strong interest in this direction. Several talks discussed finite-temperature holography, hydrodynamics  and the approach to equilibrium. David Mateos reported on numerical simulations of colliding shock waves, where he found a transition between full stopping and free crossover behaviour. Michal Heller gave a nice talk on the fundamental structure of the hydrodynamic approximation, demonstrating that for holograhic Bjorken flow, the hydrodynamic expansion has zero radius of convergence. Amos Yarom showed how contributions to the current determined by anomalies can be straightforwardly calculated from the anomaly polynomial. Toby Wiseman discussed the use of a simple estimate from a moduli space approximation to the field theory to reproduce the temperature dependence predicted by Dp-brane solutions, and argued strongly for further work on the field theory side of the correspondence at finite temperature.

Michael Gutperle reviewed higher spin gravity in three dimensions, explaining how the theory generalizes the Chern-Simons description of GR, and how black holes are described in this theory. In a companion talk, Eric Perlmutter described how this theory is related holographically to a CFT with a large affine symmetry called $hs[\lambda]$ in the large central charge limit, and reviewed some of the evidence for this duality.

There were two talks on applications of holography to condensed matter physics, with Dmitri Khveschenko discussing the subject from a condensed matter theorist's perspective, and Moshe Rozali describing a nonlinear analysis of an example of spontaneous breaking of translational symmetry in AdS. Such breakings of translational invariance arise in a number of physically interesting contexts, and pose a new challenge where the application of numerical relativity methods is playing an important role.

Pau Figueras and Sebastian Fischetti gave a tag team talk about the current status of droplet and funnel solutions in AdS, describing CFTs on black hole backgrounds in different states for the CFT. Iosif Bena gave a talk on the appearance of apparently unphysical singularities in gravitational solutions corresponding to metastable vacua, with consequences for attempts to construct de Sitter solutions in string theory by uplifting AdS vacua with antibranes. Veronika Hubeny led an informal discussion of some puzzles in the holographic calculation of entanglement entropy, considering static but not globally static cases where the original Ryu-Takayanagi and the covariant Hubeny-Rangamani-Takayanagi prescriptions disagree.

In a departure from the focus on gravity, Sheer El-Showk gave an extensive review of the application of the bootstrap in CFT in general, and discussed recent work on using the bootstrap to solve the three-dimensional Ising model.

Another theme of the meeting was instabilities, with talks by Andrzej Rostworowski, who described turbulent instability in AdS, emphasising role of resonant frequencies, and Alex Buchel, who described classes of initial data which did not show the instability, where the width of initial data seems to play a central role. Helvi Witek talked about numerical analysis of the black hole bomb instability for massive scalar fields on Kerr, and its possible astrophysical relevance. Harvey Reall in the final talk of the meeting discussed instabilities on the horizon of extreme black holes.

In informal discussion sessions, Roberto Emparan discussed his work on the large D limit of GR, showing that black hole solutions of the vacuum theory simplify in this limit, and Vojtech Pravda reviewed the current status of algebraic classification in higher dimensions.

The programme of invited talks was of exceptionally high quality, and left ample time for discussion and enjoying the delights of Benasque, from long lunches in the sunshine to hiking in the surrounding mountains. I very much look forward to the next edition in 2015. 
\vfill\eject
\section*{\centerline
{The 16th Capra meeting on radiation reaction in general relativity}}
\addcontentsline{toc}{subsubsection}{
\it Capra 16, 
by Leor Barack}
\parskip=3pt
\begin{center}
Leor Barack, University of Southampton 
\htmladdnormallink{l.barack-at-soton.ac.uk}
{mailto:l.barack@soton.ac.uk}
\end{center}
This year's Capra Meeting, the 16th in the series, took place in Dublin during the week of 15-19th July---an exceptionally warm week of Irish summer. The meeting was hosted by the School of Mathematical Sciences at University College Dublin. It was the meeting's first return to Ireland since the 2nd Capra meeting of 1999, also held at UCD. As usual, the focus was on the problem of radiation reaction and self-force in general relativity, particularly in relation to the two-body problem in the extreme-mass-ratio regime. The scientific program consisted of three days of scheduled talks, with hour-long invited reviews in the mornings followed by shorter contributed talks. In keeping with the Capra tradition, Thursday and Friday were left for loosely organized discussion sessions with plenty of free time for small-group interaction. Adrian Ottewill and his team at UCD have done a superb job in facilitating this format. 

As always, the meeting attracted a mixed audience of mathematical relativists, numerical relativists, and colleagues involved directly in gravitational-wave experiments. The program reflected this range of interests. Talks covered foundational aspects of motion in general relativity (using both  ``geometrical'' and field-theoretical approaches), relevant aspects of field dynamics in curved spacetime, and some dynamical-system aspects of EMRI systems. There was much discussion of practical schemes for actual self-force calculations in black hole spacetimes, and also much discussion of computational and numerical methods.  All talks are available in pdf format on the meeting's website at
\htmladdnormallink
{\protect {\tt{http://maths.ucd.ie/capra16/}}}
{http://maths.ucd.ie/capra16/}\\ They provide a snapshot of the activity in the field, and offer a good source of information on recent progress.  

Review talks covered advances in theory, computational method and application. Abe Harte reviewed the theory of motion of extended (but ``compact'') bodies, focusing on his non-perturbative approach. Adam Pound reviewed the perturbative approach to the equation of motion (based on matched asymptotic expansions) and described recent work to extend the formalism to second-order in the mass ratio, as well as ideas for concrete numerical calculations of second-order self-force effects. Barak Kol gave a pedagogical review of his effective-field-theory approach to the self-force in a post-Newtonian context. 
Sam Dolan surveyed the variety of techniques for concrete numerical calculations of the gravitational self-force in black hole spacetimes, focusing on the challenge of doing calculations in Kerr. Niels Warburton discussed work to compute the orbital evolution under the full effect of the gravitational self-force, covering recent results in Schwarzschild and the prospects of extending these results to Kerr. Jonathan Gair gave a comprehensive review of extreme-mass-ratio inspirals (EMRIs) as sources of gravitational waves for space-based detectors. He surveyed the history of the LISA concept and its current status, the science potential of EMRI measurements, and the data analysis challenge. 

Among the contributed talks, there were several reports of new results from self-force calculations. Abhay Shah reported on extremely high-accuracy computations (using computer algebra!) of gravitational self-force effects for circular orbits in Schwarzschild, and their successful synergy with post-Newtonian predictions. Jonathan Thornburg described results from time-domain numerical computations of the scalar-field self-force for highly eccentric equatorial orbits in Kerr, showing some new, yet unexplained features. Peter Diener talked about progress in self-force computations using full 3+1D numerics, and proposed an interesting new way for graphically representing self-force data. Patxi Ritter described work to compute the radiation reaction effect on radially falling mass particles, and Patrick Nolan discussed the electromagnetic self-force in Schwarzschild using high-order regularization parameters. Erik Forseth and Thomas Osburn presented two variants of a Lorentz-gauge calculation of the gravitational self-force along eccentric geodesic orbits in Schwarzschild, using a frequency-domain method. 

Other speakers discussed novel methodologies for self-force calculations. The motivation is twofold: to improve the computational efficiency of existing schemes, and to push forward towards calculations of the gravitational self-force for generic orbits in Kerr.
Barak Kol proposed a method for constructing the metric perturbation from an orbiting particle as a sum over partial contributions associated with certain  ``equivalent'' sources. Barry Wardell gave a frequency-domain reformulation of the standard effective-source method, in which one applies a certain regularization procedure (usually formulated in the time-domain) to the energy-momentum of the source particle. This new method is designed to enable calculations of 2nd-order self-force effects, now in progress. Seth Hopper explained how to cure the pathological behavior of the Regge-Wheeler gauge near a point particle using a local gauge deformation, and showed the application of the idea to eccentric orbits in Schwarzschild. Anna Heffernan discussed the calculation of regularization parameters for orbits in Kerr and for accelerated orbits in Schwarzschild. Cesar Merlin presented a new practical scheme for self-force calculations in Kerr, in which the force is constructed (essentially) from solutions to the Teukolsky equation, in a gauge obtained through a local deformation of the radiation-gauge perturbation. Marc Casals talked about progress in the ``matched expansions'' approach, in which the ``tail'' integral responsible for the self-force is split into a quasi-local piece, handled semi-analytically, and a far-past residual, evaluated numerically. 

The topic of transient resonances in generic Kerr inspirals received considerable attention this year. Resonances occur when the two intrinsic orbital frequencies become commensurate, leading to a loss of ergodicity, and, at least potentially, to a dramatic effect on the phase evolution. Maarten van de Meent presented a pedagogical introduction to the subject, with a summary of recent results. 
Priscilla Ca$\tilde{\rm n}$izares discussed work in progress to assess the effect of resonance crossing using an action-angle representation of the geodesic evolution. Takahiro Tanaka described the standard method for computing the orbital evolution in the adiabatic approximation (in the generic case, this also involves a method for computing the evolution of Carter's constant) and explained why it fails on a resonance. Soichiro Isoyama then described a new variant of the adiabatic evolution scheme, designed to remedy the problem and allow the computation of adiabatic evolution across resonances. 

Several speakers discussed broader theoretical aspects of the radiation reaction problem. 
Brien Nolan analyzed the fourfold singularity structure of the Green's function in Schwarzschild spacetime through an asymptotic analysis at high multipole numbers, and explained the relevance to the self-force problem. Chad Galley showed how his effective-field-theory approach can be used to compute the gravitational self-force in the ultra-relativistic limit to high order in perturbation theory. He also speculated on the interesting problem of finding the self-force acting on photons. 
Dmitry Gal'tsov explained how classical gravitational radiation damping in trans-Planckian scattering may provide yet another mechanism of ``classicalization'' in multidimensional quantum gravity (in addition to black hole creation), offering an escape from nonunitarity in scenarios where black holes aren't created. 
Peter Zimmerman discussed the contribution to the gravitational self-force from non-gravitational fields that couple to gravity, and Shahar Hadar presented advances in the formulation of dissipative effects within the effective field theory approach to the post-Newtonian dynamics in gravitating binaries.  

The (quasi)formal part of the meeting ended with a customary windup talk by Bernard Whiting (despite his protest for being typecast for the job!). After reflecting on progress and prospects, Bernard commented observantly about ``social'' trends in the Capra community, especially the increasing role played by young contributors---a very welcome trend indeed. 

Wednesday afternoon provided an opportunity to enjoy the fine weather, with an excursion to the seaside town of Howth (rhymes with ``both'')---a semirural suburb of Dublin beautifully located on a headland peninsula at the north of Dublin bay. The trip included a scenic cliff walk around Howth Head and concluded with a fine dinner at Howth harbour. The sweeping vistas from atop the headland provided a perfect backdrop for reflecting on what had been a well-organized, enjoyable and engaging Capra meeting. 
\vfill\eject

\section*{\centerline
{Workshop on Reduced Order Modeling in General Relativity}}
\addcontentsline{toc}{subsubsection}{
\it Reduced Order Modeling in General Relativity, 
by Michele Vallisneri}
\parskip=3pt
\begin{center}
Michele Vallisneri, Jet Propulsion Laboratory
\htmladdnormallink{vallis-at-caltech.edu}
{mailto:vallis@caltech.edu}
\end{center}

The first workshop on \textbf{Reduced Order Modeling in General
Relativity} was held at Caltech on June 6 and 7, 2013. The workshop was
organized by Manuel Tiglio, Jan Hesthaven, Scott Field, and Chad Galley,
and it brought together experts from the communities of applied
mathematics and gravitational waves/numerical relativity, to discuss new
opportunities for interactions and collaborations using the recently
developed tools of reduced-order models (ROMs) and bases. Broadly
speaking, ROMs are low-dimensional mathematical representations of
complex, high-dimensional numerical problems (such as the solution of
PDEs on a grid as a function of a number of parameters), which provide
remarkable accuracy and yet are very computationally inexpensive to
evaluate. They may however be rather expensive to set up (by
accumulating a large number of ``system responses'' to ``input
excitations''), but this \emph{offline} stage needs to be performed only
once, while its subsequent \emph{online} use is generally very
efficient.

The slides from all the talks given at the workshop can be found at
\href{}{http://www.tapir.caltech.edu/\ensuremath{\sim}rom-gr/participants.html}.
On the first day, \textbf{Alan Weinstein} (Caltech) introduced
gravitational waves, the efforts to detect them, and their possible
sources. He emphasized how different detection techniques are tailored
to different signals, focusing on matched-filtering searches for compact
binary coalescences, discussing the construction of signal templates and
the placement of template banks, and pointing out that the reduction of
large banks to a smaller number of ``independent'' shapes (as done in
\emph{gstlal}) is in effect a ROM technique. \textbf{Mark Scheel}
(Caltech) gave an overview of numerical relativity and the problem of
binary black-hole mergers, introducing different formulations and
sketching out the implementation details of
\href{http://www.black-holes.org/SpEC.html}{SpEC}, the
Caltech/Cornell/CITA/\ldots{} spectral code. He highlighted the current,
consistently expanding capabilities of modern codes (high spins, higher
mass ratios, precessing systems, accurate waveforms,
\href{http://arxiv.org/abs/1304.6077}{large catalogs}), and their
challenges (covering large parameter spaces, extending runs to many
orbits, achieving accurate initial data, fitting to analytical models,
reducing computational cost).

\textbf{Jan Hesthaven} (Brown) introduced the general technique of
\emph{reduced bases}, whereby the set of solutions to a parametric
equation is approximated as a sum over a small, optimally selected
basis. Then the solution for new parameter values can be recovered
accurately with little cost by, in effect, solving a reduced-complexity
version of the original equations. (E.g., a supercomputing
finite-element analysis of a bridge may be
\href{http://people.seas.harvard.edu/~dknezevic/research.html}{reproduced
on a smartphone}.) Clearly the careful construction of the basis is
paramount: while exact, exhaustive approaches are too expensive,
Hesthaven discussed how the basis can be built practically using
effective error estimates and \emph{greedy algorithms} (which, roughly
speaking, approximate global maxima by consistently seeking local
maximizations). As an example, Hesthaven showed an impressive
\href{http://arxiv.org/abs/1101.3765}{demonstration} of the
reduced-basis representation of binary---neutron-star inspiral
waveforms. The most serious remaining bottleneck is the sampling of
high-dimensional parameter spaces, which can be mitigated by isolating
the most influential combinations of parameters (using, e.g.,
\emph{ANOVA expansions}). \textbf{Peter Binev} (South Carolina)
presented a mathematical treatment of greedy strategies, studying the
cases of binary-tree approximation (which arises, e.g., in image
compression) and the reduced-basis solution of parameter-dependent PDEs.
He focused on possible variants of the strategies, their convergence
rate, and robustness.

\textbf{Don Estep} (Colorado State) discussed the broad topic of
\emph{uncertainty quantification} --- the characterization of the
effects of errors and uncertainties as they propagate through a physical
model, in either direction: \emph{forward sensitivity analysis}
characterizes the dependence of model predictions on parameter
uncertainties and numerical/simulation errors; \emph{inverse sensitivity
analysis} characterizes the uncertainty of inferred system parameters
deriving from noisy observations. Dimensional reduction is achieved
automatically in techniques that sample parameter space adaptively,
adding simulations/observations in the regions and dimensions that
contribute the strongest effects. \textbf{Wojciech Czaja} (Maryland)
examined mathematical techniques known as \emph{kernel eigenmap
methods}, which lie at the intersection of harmonic analysis and machine
learning. Roughly speaking, these techniques are relatives of the
familiar \emph{principal component analysis}; they express the
similarities of data vectors in terms of a quadratic \emph{kernel
function}, then diagonalize the kernel and reduce its dimensionality by
(again, roughly speaking) selecting a few diagonal elements. These
methods give impressive results in problems of classification and
detection, and can be accelerated using random projections (more below).

On the second day, \textbf{Yanbei Chen} (Caltech) presented an
exhaustive overview of the semianalytical treatment of compact-binary
inspirals, discussing the building blocks of the post-Newtonian
formalism, the various possible \emph{resummations} of
adiabatic-inspiral equations, the \emph{effective-one-body} approach,
the theory of black-hole perturbations and quasinormal modes, and the
problem of matching analytical and numerical waveforms. Chen emphasized
that semianalytical models can provide computational efficiency and
physical insight, and that they can be calibrated efficiently using
numerical results---at least for nonspinning binaries, but work is in
progress for the much more complex spinning systems. \textbf{Michele
Vallisneri} (JPL, the author of this report) discussed the application
of Markov-chain Monte Carlo techniques to the problem of inferring
physical parameters from gravitational-wave observations. He recounted
the intriguing history of Monte Carlo methods and of the Metropolis
algorithm, and argued that physicists may like these methods so much
because they solve a mathematical problem by harnessing the power and
insight of statistical physics. Vallisneri explained why Markov-chain
methods converge to sample their target probability distribution; he
analyzed their practical shortcomings and possible remedies; and he
discussed two broadly used, modern algorithms
(\href{http://ccpforge.cse.rl.ac.uk/gf/project/multinest/}{nested
sampling} and \href{http://dan.iel.fm/emcee}{affine--invariant
sampling}).

\textbf{Justin Romberg} (Georgia Tech.) gave an overview of the
(recently) acclaimed field of
\href{http://dsp.rice.edu/cs}{\emph{compressive sensing}}, whereby a
signal that is known to be (in some sense) \emph{sparse} can be acquired
and reconstructed using a very small set of observations---typically,
the projections of the data on an incoherent, quasirandom basis. The
crucial mathematical observation is that a \emph{random} $m \times n$
matrix $\Phi$ (with $m \ll n$) will keep two sparse $n$-dimensional
vectors $x_1$ and $x_2$ \emph{separated}, in the sense that
$||\Phi (x)(x_1 - x_2)|| \simeq ||x_1 - x_2||$. Practical applications
include the reconstruction and interpretation of sequences of
multispectral, high-resolution images taken from satellites, the
acceleration of MRI data acquisition, and more. Romberg's more
mathematical description of the field was that ``underdetermined systems
of linear equations can be meaningfully inverted if they have
\emph{structured} solutions.'' Random projections can also be used to
provide surprisingly efficient reduced bases for forward modeling
problems. (From the standpoint of gravitational-wave detection, we
should temper expectations by noting that compressed sensing cannot be
used in low signal-to-noise--ratio situations where every available
sample must be used to beat down the noise.) \textbf{Akil Narayan}
(UMass. Dartmouth) discussed the role of interpolation (of approximated
solutions across parameter values) in ROM, covering both non-adaptive
techniques (smart, greedy generalizations of the classical selection of
optimal interpolation nodes) and adaptive techniques (which place
interpolation nodes according to the structure of the solutions). Among
the latter, the \emph{empirical interpolation method} is especially
suited to reduced-basis methods: it provides interpolation coefficients,
which are needed to approximate any solution as a sum over the basis, in
terms only of the value of the solution at a small set of
\emph{greedily} chosen coordinates.

\textbf{Gianluigi Rozza} (SISSA) reviewed the application of
reduced-basis methods to problems in hydrodynamics, and specifically to
the modeling and simulation of viscous flows in the human cardiovascular
system. In this context, ROM is used for data assimilation (e.g.,
reconstructing blood flows and arterial configurations) and control
problems (e.g., optimizing the design of cardiovascular devices given
the data from the patient). Last, \textbf{David Knezevic} (Harvard, the
coauthor of the
\href{http://people.seas.harvard.edu/~dknezevic/research.html}{smartphone
bridge applet}) discussed the application of reduced-basis methods to
real-time finite-element applications: control and acquisition in
LabVIEW, experimental parameter estimation, and \emph{in situ}
validation of samples. He outlined \emph{component-based} reduced-basis
methods, which combine reduced bases with domain decomposition to build,
for instance, a library of acoustic components with ROM descriptions
that can be combined together.

All in all, this workshop left this writer, a relativist and data
analyst, in awe of the sophistication and ingenuity of applied
mathematicians, who have figured out how to leverage specific knowledge
about a system to circumvent seemingly invulnerable performance limits
(such as posed by the Nyquist theorem, or by well-posedness of
linear-algebra problems). As for the application of ROM to general
relativity, the direct reduction of the complex and very nonlinear
Einstein equations appears for the moment to be a distant goal; indeed,
the algorithms developed in the applied-math and engineering contexts
can only deal with mild nonlinearity. Instead, much of the effort in our
community has gone toward the reduced-basis approximation of waveforms,
and it recently reached a culmination with the
Field/Galley/Hesthaven/Kaye/Tiglio formalism of
\href{http://arxiv.org/abs/1308.3565}{surrogate models}.

The idea there is to accurately approximate \emph{any} waveform in a
parametrized family as a sum over a precomputed reduced basis, using
interpolation coefficients that are functions of the system parameters
derived from the basis itself. The \emph{online} use of such a model for
matched filtering or parameter estimation promises staggering
performance gains---especially so for waveform families that are
expensive to compute, such as EOB or hybrid compact-binary waveforms,
and that depend on a small number of parameters. A surrogate model is
obtained in three steps, which are expensive and performed offline:
first, an initial dense bank is \emph{greedily} distilled to a much
sparser reduced basis; second, \emph{greedy} empirical interpolation is
used to identify times such that the reduced-basis decomposition of any
waveform can be approximated from its value at those times only; third,
the reduced-basis waveforms are used to compute fitting functions for
the amplitude and phase of the waveforms at the interpolation times, as
a function of the system parameters. If this is confusing, go read the
paper; but truly it seems (as professed by one of the workshop
presenters, channeling Gordon Gekko) that sometimes \emph{greed is
good}.

\end{document}